# SIMULATING OUR COSMOLOGICAL NEIGHBORHOOD: MOCK CATALOGS FOR VELOCITY ANALYSIS

Tsafrir Kolatt[1,2], Avishai Dekel[2]
Galit Ganon[2], and Jeffrey A. Willick[3]

[1] Harvard-Smithsonian Center for Astrophysics, 60 Garden St., Cambridge MA 02138 USA.

[2] Racah Institute of Physics, The Hebrew University, Jerusalem 91904 Israel.

[3] Carnegie Observatories, 813 Santa Barbara St., Pasadena CA 91101 USA.

# ABSTRACT

We describe the construction of an N-body simulation that mimics the true velocity and mass-density fields in a box of side 256 $h^{-1}$Mpc about the Local Group, and the production of mock catalogs that mimic in detail current catalogs of redshifts and peculiar velocities. Our main purpose is to provide a tool for developing and testing reconstruction methods, but the different components of the method can be used on their own in other applications.

The initial conditions in the present application are based on the $IRAS$ 1.2Jy redshift survey, assuming that galaxies trace mass and $\Omega = 1$. A density field smoothed with a Gaussian of radius 5 $h^{-1}$Mpc is recovered from the redshift survey, using quasi-linear theory and a power-preserving filter. The corresponding potential field is traced back to the linear regime using the Zel'dovich-Bernoulli equation. Small-scale power is added by means of constrained realization to mimic fluctuations on galactic scales. The gravitating system is evolved forward in time with a PM code of 2 $h^{-1}$Mpc resolution, and stopped when $\sigma_8 = 0.7$. The result reproduces the real dynamical structures on large scales and the statistical properties of the structure down to galactic scales.

"Galaxies" are identified via a linear biasing scheme ($b = 1.35$) and they are divided into "spirals" and "ellipticals" to obey Dressler's morphology-density relation. The galaxies are assigned internal-velocity parameters ($\eta$) and absolute magnitudes scattered about an assumed mean Tully-Fisher relation. They are then "observed" as magnitude-limited samples, trying to mimic the selection criteria of the data sets constituting the *Mark III* catalog of peculiar velocities. Artificial $IRAS$ 1.2Jy redshift surveys are also compiled. The simulations and mock catalogs will be made available electronically as benchmarks for testing reconstruction methods.

# 1. INTRODUCTION

The analysis of large-scale peculiar velocities has become an active field of research with important cosmological implications. Radial velocities provide direct dynamical constraints on the gravitating fields of velocity and mass-density fluctuations, and they address the basic cosmological parameters, the initial density fluctuations, and the nature of the dark matter. Combined with redshift surveys they also constrain the "biasing" process of galaxy formation. Recent reviews of this field are provided by Dekel (1994) and by Strauss & Willick (1995). The most comprehensive data set today is the *Mark III* catalog (Willick *et al.* 1995, WI; 1996a, WII; 1996b, WIII), based on ∼3500 galaxies. Methods for recovering the dynamical fields from the observations have been developed, led by the POTENT algorithm (Bertschinger & Dekel 1989; Dekel, Bertschinger & Faber 1990, DBF).

Peculiar velocity measurements are noisy and relatively poorly sampled as a result of the elaborate observations involved and the limited accuracy of distance indicators such as the Tully-Fisher and $D_n$–$\sigma$ relations. These limitations introduce severe random and systematic errors into the recovered dynamical fields. The main effort in this field is therefore devoted to developing reconstruction methods that minimize these errors. The development and testing of these methods rely on artificial catalogs in which the true velocity and mass-density fields are known, and which properly simulate all the important sources of error. The production of such mock catalogs is complicated by the fact that the final errors depend not only on the observational selection process and the random distance errors, but also on the underlying fields themselves. For example, non-uniformity in the sampling introduces a "sampling-gradient bias" (DBF) which is a strong function of the gradients in the velocity field on scales comparable to the desired smoothing scale (e.g. it vanishes for a constant velocity field). As another example, the "inhomogeneous Malmquist bias" (DBF; Willick 1995) depends on the underlying density of galaxies from which the catalogs were selected. The simulated catalogs should therefore be drawn from a dynamical simulation that resembles our actual cosmological neighborhood.

This paper describes the construction of such simulations and galaxy catalogs. In the present case, the mass distribution is made consistent with the IRAS 1.2Jy redshift survey (Fisher *et al.* 1995) assuming that IRAS galaxies trace mass and $\Omega = 1$. Galaxies are identified according to their observed statistical properties, and they are "observed" in a way that mimics in detail the *Mark III* catalog of peculiar velocities. The resultant mock catalogs are meant to serve as standard benchmarks for testing the various versions of the POTENT algorithm (Dekel *et al.* 1996) as well as alternative reconstruction methods (*e.g.* , Nusser & Dekel 1996). Artificial IRAS 1.2Jy redshift surveys are also "observed." They will help test methods of reconstruction from redshift surveys, as well as comparisons of peculiar velocity and redshift data aimed at determining $\Omega$ and the galaxy-biasing scheme.

In §2 we describe the dynamical simulation, starting from the IRAS survey and ending in a full N-body simulation evolved to the present. In §3 we discuss the identification of galaxies and the assignment of observable quantities to them. In §4 we explain how we "observe" the simulated universe and produce mock *Mark III* and IRAS catalogs. In §5 we evaluate the success of the procedure, comment on its present limitations, suggest improvements, and discuss the use of the mock catalogs.

## 2. THE DYNAMICAL SIMULATION

We wish to simulate a non-linear gravitating system whose mass distribution and velocity field resemble as closely as possible our real cosmological neighborhood on scales ranging from galactic scales to several hundred megaparsecs. On the linear and quasi-linear scales, $\gtrsim 10$ $h^{-1}$Mpc, we make the simulation resemble the actual structure as traced by galaxies, or as derived by POTENT from peculiar velocities. On smaller, non-linear scales we fill in a random realization of structure with statistical properties that resemble those of the true velocity and density fields as traced by galaxies, while obeying the constraints imposed by the actual large-scale structure.

The procedure consists of the following five steps:

1. Adopt a quasi-linear density field for the present structure. Here we derive the present gravitating density, velocity and potential fields, Gaussian smoothed with radius 5 $h^{-1}$Mpc (G5), from the IRAS 1.2Jy redshift survey. We use a power-preserving filter (PPF) to reduce shot noise, and assume that IRAS galaxies trace mass and $\Omega = 1$.

2. Trace the structure back to the linear initial conditions. We do it by integrating back the Eulerian Zeldovich-Bernoulli equation.

3. Force the initial one-point probability distribution function (PDF) to be Gaussian, to remove small artificial effects introduced in step 1.

4. Fill in a Gaussian random realization of small-scale power constrained by the smoothed structure obtained in step 2.

5. Represent the initial density fluctuations as an N-body system and evolve it forward in time using a Particle-Mesh (PM) N-body code until it resembles the present day structure over the simulated range of scales.

### 2.1. The Present Quasi-linear Density Field from IRAS

As an approximation for the present quasi-linear mass density field in the local universe we adopt a smoothed density field derived by A. Yahil and M. Strauss (private communication) from the IRAS 1.2Jy redshift survey of galaxies (Fisher *et al.* 1995). The density field in real space, along with the associated velocity and potential fields, are obtained from the redshift distribution as a self-consistent, quasi-linear gravitating solution via an iterative procedure (see Yahil *et al.* 1991; Strauss & Willick 1995; Strauss *et al.* 1995). It assumes that the *IRAS* galaxies trace mass up to a known selection function, and that $\Omega = 1$.

To evaluate a likely estimate of the true density field from the data, which are contaminated by radially-increasing shot-noise due to sparse sampling, we filter the data in shells in the spirit of the Wiener filter (see Press *et al.* 1994). The standard Wiener filter provides the least-squares approximation to the true signal given the noisy data and an assumed prior power-spectrum $P(k)$. The Wiener filter is $F(k) = P(k)/(P(k) + N^2)$, where $N^2$ is the typical noise in the given shell. However, this filter attenuates the variance of the density field in regions where the noise dominates, which effectively acts like variable smoothing. This is an undesirable feature when the system is analyzed as a gravitating system, and does not give a uniform representation of the local universe. A. Yahil (private communication) has proposed a slight modification into a filter that recovers a signal of constant variance,

$$F_{ppf}(k) = \sqrt{\frac{P(k)}{P(k) + N^2}}. \tag{1}$$

The recovered signal is no longer a least-square approximation, but can be shown to deviate little from it. Note that the higher moments of skewness etc. are not necessarily preserved, and we will have to correct for that (§2.3).

We adopt as a prior the power-spectrum derived from the *IRAS* 1.2Jy survey itself (Fisher *et al.* 1993)

$$P(k) = \frac{Ak}{\alpha + k^{4-\gamma}}, \tag{2}$$

with $\alpha = 7.17 \times 10^{-4}$ and $\gamma = 1.66$. The variance of the shot-noise at distance $r$ from the Local Group (LG) is $N^2 \approx d^3/V_w$, where $d$ is the mean separation between *IRAS* galaxies at $r$, and $V_w$ is the effective volume of the smoothing window, here a 3D G5. The mean separation in the range $40 - 80$ h$^{-1}$Mpc is approximated by $d \approx 4.5 \times 10^{r/133}$ h$^{-1}$Mpc.

The recovered, G5-smoothed density field is obtained on a cubic grid of spacing 2 h$^{-1}$Mpc inside a cubic box of side 256 h$^{-1}$Mpc centered on the LG. Periodic boundary conditions are imposed by zero padding (apodizing with a top-hat sphere.) Figure 1 shows a map of this density field in the Supergalactic plane. The rms of the G5 density fluctuation turns out to be $\sigma = 0.54$, while in top-hat spheres of radius 8 h$^{-1}$Mpc (TH8) it is $\sigma_8 = 0.64$.

## 2.2. Back to the Linear Regime

The G5 density field recovered at the present time (§2.1) is not linear, with $\delta \sim 10$ at the high peaks and $\delta \sim -0.7$ in the deep voids. A constrained realization of small-scale power is not applicable at this stage because the PDF is severely non-Gaussian. We wish to first trace the fluctuations back in time to the linear regime, where, presumably, the field was Gaussian (see Nusser, Dekel & Yahil 1995).

Naive backwards integration of the equations of gravitational instability (GI) would in general fail to recover the special initial state of small fluctuations, as noise would be amplified by the "decaying" modes into spurious initial fluctuations. This problem can be solved either by applying the principle of least action (Peebles 1989) or by eliminating the decaying modes (see a discussion in Dekel 1994, §7.2).

We adopt here the latter approach and apply the "time machine" of Nusser and Dekel (1992). The Zel'dovich approximation for GI (Zel'dovich 1970), which is restricted to the growing mode, has been translated to Eulerian space. When applied to a potential flow such as our quasi-linear gravitating system, it yields the Zel'dovich-Bernoulli equation for the velocity potential,

$$\frac{\partial \varphi_v}{\partial t} - \frac{\dot{D}}{2}(\boldsymbol{\nabla}\varphi_v)^2 = 0, \qquad (3)$$

where $\varphi_v(\boldsymbol{x},t)$ is in units of $a^2\dot{D}$, $a(t)$ is the universal expansion factor, and $D(t)$ is the growing-mode solution of GI (e.g. Peebles 1993). The velocity potential is related to the gravitational potential in the quasi-linear regime (e.g. in the Zel'dovich approximation) by $\varphi_v = 2f(\Omega)/(3H\Omega)\varphi_g$, where $f(\Omega) \approx \Omega^{0.6}$. The Zel'dovich-Bernoulli equation can easily be integrated backwards in time with a guaranteed convergence to uniformity at early times. The initial velocity and density fields are then derived from the initial potential using linear theory. The initial fields, up to a scaling factor, are applicable at any desired time in the linear regime.

Figure 2 shows the linear density field in the Supergalactic plane, arbitrarily normalized to $\sigma_8 = 1$.

## 2.3. Gaussianization

The initial PDF of the obtained field shows slight deviations from a Gaussian distribution, which vary as a function of distance. Apart from shot-noise and cosmic scatter in the *IRAS* data, these deviations were verified by N-body simulations to be mostly an artifact of the PPF filtering, which does not preserve the high moments of the PDF, and it does it to a degree that varies as a function of noise, and thus distance (Figure 3a).

We impose Gaussianity by a rank-preserving procedure in shells. In order to minimize possible cosmic scatter effects we use three thick shells of radii: $r < 70$ h$^{-1}$Mpc, $70 <$ $r < 100$ h$^{-1}$Mpc, $r > 100$ h$^{-1}$Mpc. In each shell, $i$, we compute the rank-preserving transformation $\delta \to \delta_G^i$, which would have corrected the PDF in that shell alone into a Gaussian. In order to impose continuity across the shell boundaries, we actually transform the $\delta$ value at a point $\boldsymbol{r}$ into an average of the $\delta_G^i$ values, weighted inversely by the difference between $r$ and the mean radius of shell $i$.

This procedure yielded a reasonable approximation to a Gaussian PDF over the whole box (Figure 3b), which enables a rigorous implementation of the constrained realization technique (§2.4). The Gaussianized, smoothed density field is shown in Figure 4.

### 2.4. Constrained Realizations of Small-Scale Power

We want the mock catalog to represent the dynamics of groups of galaxies, as the peculiar velocities of galaxies and their local clustering are expected to strongly affect the biases entering the velocity analysis. This dictates spatial and mass resolution of at least $\sim 2$ h$^{-1}$Mpc for the initial conditions and for the N-body simulation. The fluctuations on small scales are filled in as a Gaussian random realization of a prior power spectrum, constrained by the G5-smoothed density field recovered in §2.3, using the method of Hoffman & Riback (1991, 1992) as implemented by Ganon & Hoffman (1993).

The first step in this procedure is to compute the most likely mean field of density fluctuations $\delta_{mf}(\boldsymbol{r})$, which obeys the discretely sampled constraints, $c_j$, with their associated errors, under the assumption of the prior model. This mean field is given by

$$\delta_{mf}(\boldsymbol{r}) = \xi_i(\boldsymbol{r}) \xi_{ij}^{-1} c_j. \tag{4}$$

The matrix $\xi_{ij}$ is the auto-correlation matrix of the field at the points of constraints, which consists of the model auto-correlation plus the auto-correlation of experimental errors. The vector $\xi_i(\boldsymbol{r})$ is the cross-correlation between the model density at the position of the constraint $c_i$ and the model density at $\boldsymbol{r}$.

The second step is to generate a Gaussian random realization field of the model power spectrum about a zero mean, $\delta'(\boldsymbol{r})$. The constrained random realization is then taken to be

$$\delta(\boldsymbol{r}) = \delta'(\boldsymbol{r}) - \xi_i(\boldsymbol{r})\xi_{ij}^{-1}c'_j + \delta_{mf}(\boldsymbol{r}), \tag{5}$$

where $c'_j$ are the values of $\delta'$ at the points where the constraints $c_j$ are given. The computation involves fast Fourier transform, matrix inversion and matrix multiplication.

We take the prior to be the power spectrum derived from the *IRAS* 1.2Jy survey (equation 2 appropriately normalized) and smoothed with a Gaussian of radius 2 h$^{-1}$Mpc (G2). The constraints from the G5 smoothed density field are taken at cubic grid points covering the 256 h$^{-1}$Mpc box with spacing of 12 h$^{-1}$Mpc. The smoothed power spectrum of the constraints is approximated by equation (2) times a filter of the form

$$f(k) = \begin{cases} \exp(\alpha k^2 + \beta k + \gamma), & k \geq 0.107(\text{ h}^{-1}\text{Mpc})^{-1} \\ 1 & \text{otherwise.} \end{cases} \quad (6)$$

with the fitting parameters $\alpha = 1.96$, $\beta = -13.61$ and $\gamma = 1.44$.

Figure 5 shows the density field of the actual initial conditions of the simulation.

### 2.5 N-body Simulation Forward in Time

Using the density and velocity fields obtained in the previous step, we construct a particle distribution as an initial condition for an $N$-body simulation. We use $128^3$ equal-mass particles, appropriately perturbed from a cubic grid, and run a Particle-Mesh code (Bertschinger & Gelb 1991) with $128^3$ grid points inside the cubic box of side 256 h$^{-1}$Mpc. With $\Omega = 1$, the mass per particle is $\approx 2.1 \times 10^{12} h^{-1} M_\odot$. The simulation was stopped when $\sigma_8$, the rms fluctuation of mass density in top-hat spheres of radius 8 h$^{-1}$Mpc, reached the value 0.7 (see §3.1.)

Figure 6 shows the final distribution of mass particles in a slice about the Supergalactic plane, and the corresponding G12-smoothed mass density field. Figure 7 shows the mass density field smoothed by a G5 window. Apart from the slightly different normalization, the final mass distribution of Figure 7 indeed resembles the smoothed density derived from *IRAS* galaxies (Figure 1).

The observational constraints require a low velocity dispersion (see §3.1 below). The simulations are designed to obey this constraint by the G2 smoothing of the initial conditions (§2.4), followed by the $\approx 3$ h$^{-1}$Mpc force resolution of the PM code, and combined with the early halt of the simulation at $\sigma_8 = 0.7$. Such a low value of $\sigma_8$ is also required in order to explain the abundance of rich clusters (White, Efstathiou, & Frenk 1993). The final dispersion of particle peculiar velocities about the bulk flow within 10 h$^{-1}$Mpc spheres turns out to be 340 km s$^{-1}$.

It is worth noting that no special attempt has been made to artificially mimic a cold flow in the vicsinity of the "Local Group". The Mach number in a sphere of radius 20 h$^{-1}$Mpc about the origin is 0.53, comparable to the *rms* value across the simulation, of 0.58. The slight excess of velocity dispersion near the "Local Group" can be mostly attributed to the fact that the Virgo cluster is a spiral-rich cluster and is therefore prominent in the *IRAS* density field.

# 3. IDENTIFYING GALAXIES IN THE SIMULATION

Given the N-body particles, each roughly representing a galactic mass, we wish to first choose a volume limited sample of galaxies ("$g$") according to an assumed "biasing" recipe. We then divide them into E's (ellipticals and S0's, marked "$e$") and S's (spirals and irregulars, marked "$s$"), in such a way as to obey the observed morphology-density relation. Finally, we assign them observable quantities such that we can later mimic the sample selection and the distance estimation.

### 3.1. Volume-Limited, Biased Galaxy Distribution

The mean number density of galaxies is chosen to obey the following constraints:

1. The desired mean density of S galaxies, based on the mean in the $IRAS$ 1.2Jy redshift survey had the selection function been unity everywhere, is $\bar{n}_s = 0.057(\,\text{h}^{-1}\text{Mpc})^{-3}$ (Fisher *et al.* 1994, Yahil *et al.* 1991).

2. The desired total fraction of S's is $f_s \approx 0.8$ (de Vaucouleurs *et al.* 1991, Lauberts *et al.* 1989).

3. The mean density in the simulation is $\bar{n} = 0.125(\,\text{h}^{-1}\text{Mpc})^{-3}$ (§2).

¿From constraints (1) and (2) we deduce a desired mean galaxy density of

$$\bar{n}_g = \bar{n}_s/f_s \approx 0.071 \ . \tag{7}$$

The global ratio of galaxies to particles is thus $\bar{n}_g/\bar{n} = 0.57$. We consider all particles as potential candidates, and choose galaxies such that the mean galaxy density is as desired.

The simulation is stopped and galaxy identification is made such that the following constraints are roughly obeyed:

4. The dispersion of galaxy density fluctuations in top-hat spheres of radius 8 $\text{h}^{-1}$Mpc is, based on optical catalogs, $\sigma_{8g} = 0.95 \pm 0.05$ (de Lapparent *et al.* 1988).

5. For S galaxies, based on the $IRAS$ 1.2Jy redshift survey, $\sigma_{8s} = 0.65 \pm 0.05$ (Strauss *et al.* 1995)

6. The 1D dispersion of pair velocities at separation 4 $\text{h}^{-1}$Mpc for spirals is $\sigma_v \sim 150 - 200$ km s$^{-1}$. In the $IRAS$ 1.2Jy redshift survey it is estimated to be $\approx 200$ km s$^{-1}$ (Fisher *et al.* 1994). In the SSRS+CfA2 optical surveys, excluding clusters, it is $\approx 190$ km s$^{-1}$ (Marzke *et al.* 1995). A comparison of the Mark III data to the velocities predicted from the IRAS 1.2Jy redshift survey yields $\sigma_v \approx 150$ km s$^{-1}$ (Willick *et al.* in preparation; already reported in Strauss & Willick 1995). A similarly low value is obtained from a recent Optical Redshift Survey (Strauss & Ostriker in preparation).

7. Based on simulations of galaxy formation (Cen & Ostriker 1993), the linear biasing factor of galaxy density fluctuations smoothed on scales of several megaparsecs is $b = \delta_g/\delta \sim 1.4$.

The desired low velocity dispersion (6) led us to stop the simulation when the mass fluctuations were $\sigma_8 = 0.7$. With constraint (4), the linear biasing factor is

$$b = \sigma_{8g}/\sigma_8 \approx 1.35 \qquad (8)$$

as desired in (7). Based on constraint (5), the S galaxies are effectively unbiased on scales of several megaparsecs and beyond.

For choosing galaxies at random from the particles we evaluate the conditional probability for a particle to be a galaxy, which we identify with the ratio of unsmoothed number densities: $P(g|p) = n_g/n$. In practice, we replace these densities by the corresponding G5-smoothed densities, $\tilde{n}_g$ and $\tilde{n}$. This is roughly equivalent to TH8 smoothing. Since the window smoothing is a convolution that operates on $n$ and $n_g$ in a similar way, the ratios are the same: $\tilde{n}_g/\tilde{n} = n_g/n$.

We adopt a deterministic linear *biasing* model for the G5 smoothed density fluctuations,

$$\tilde{\delta}_g(\boldsymbol{r}) = b\,\tilde{\delta}(\boldsymbol{r}). \qquad (9)$$

The resulting $\tilde{\delta}_g$ is properly $\geq -1$ only for $\tilde{\delta} \geq -b^{-1}$. For $b = 1.35$ and G5 smoothing, this inequality is invalid only for a very small fraction of one in $\sim 10^5$ grid points, for which we can adopt $P(g|p) = 0$ with negligible effect on the moments of $\tilde{\delta}_g$. This scheme defines a proper density fluctuation field because $\langle \tilde{\delta}_g \rangle = b\langle \tilde{\delta} \rangle = 0$ as required by definition.

Finally, from the definition $\tilde{n}/\bar{n} = 1 + \tilde{\delta}$, we obtain:

$$P(g|p) = \frac{\tilde{n}_g}{\tilde{n}} = \frac{\bar{n}_g}{\bar{n}}\frac{1+b\tilde{\delta}}{1+\tilde{\delta}}, \qquad (10)$$

where $\tilde{\delta}$ is the G5 smoothed density fluctuation of mass at the particle position. This probability function is properly bounded by $\leq 1$ for any $b < \bar{n}/\bar{n}_g = 1.75$, and in particular for the desired $b = 1.35$. $P$ is appropriately positive for $\tilde{\delta} \geq -b^{-1} = -0.75$, and $P = 0$ where $\tilde{\delta}$ is smaller.

### 3.2. Assigning Galaxy Types

The galaxies are divided into S's and E's based on an assumed morphology-density relation. Each galaxy is randomly labeled S with probability $P(s|g,\tilde{\delta}_g)$, and E with probability $P(e|g,\tilde{\delta}_g) = 1 - P(s|g,\tilde{\delta}_g)$.

Since we assumed that the desired S galaxies should be relatively unbiased on scales of several megaparsecs ($\sigma_{8s} \approx \sigma_8 \approx 0.7$), we identify S's at low $\tilde{\delta}_g$ by roughly redoing the linear biasing of Eq. (10), namely

$$P(s|g,\tilde{\delta}_g) = \begin{cases} f'_s \left( \frac{1+b'^{-1}\tilde{\delta}_g}{1+\tilde{\delta}_g} \right), & \tilde{\delta}_{g_1} \leq \tilde{\delta}_g < \tilde{\delta}_{g_2} \\ 1, & \tilde{\delta} \leq \tilde{\delta}_{g_1} \end{cases} \quad (11)$$

where $\tilde{\delta}_{g_1} = -(1 - f'_s)/(1 - f'_s b'^{-1})$. We allow slight adjustments in $f'_s$ and $b'$ compared to the values of $f_s$ and $b$ assumed before, in order to fit better the desired dispersions at TH8 smoothing, and adopt after trial and error $f'_s = 0.85$ and $b' = 1.5$. (The quantities $f'_s$ and $b'$, which slightly differ from $f_s$ and $b$, are used only in the context of Eq. 11.)

At high densities, $\tilde{\delta}_g \geq \tilde{\delta}_{g_2}$, we try to mimic the morphology-density relation by Dressler (1980). The transition point was chosen such that the transition is continuous, $\tilde{\delta}_{g_2} = 2.4$, where $P(s|g,\tilde{\delta}_g) = 0.65$.

Dressler defines the local galaxy density, $n_d$, within the sphere encompassing the $N = 10$ nearest neighbors in his sample. A good fit to the data in Figure 4 of Dressler (1980) is provided by

$$P(e|g,n_d) = 0.00417 \, (\log n_d)^2 + 0.125 \, (\log n_d) + 0.345, \quad (12)$$

where the density is measured in $(h^{-1}\text{Mpc})^{-3}$.

In order to translate this into a conditional probability given $\delta_g$, which is calculable in our simulation, we derive in Appendix A the correspondence

$$n_d(n_g) = \frac{\bar{n}_d}{\bar{n}_g} n_g f_{pm}(n_g), \quad (13)$$

with $\bar{n}_d/\bar{n}_g = 0.145$. The factor $f_{pm}$, which corrects for the limited grid resolution in the PM simulation, is approximated by

$$\log f_{pm}(\tilde{n}_g) = \begin{cases} 0, & n_g \leq n_g^0 \\ A[(n_g/n_g^0) - 1]^\gamma, & n_g > n_g^0, \end{cases} \quad (14)$$

with $n_g^0 = 29.2$, $A = 0.5$, and $\gamma = 0.7$. The local density in this case should be measured by $n_g = n_{70} \equiv 70/V_{70}$, where $V_N$ is the volume occupied by the nearest $N$ galaxies.

Following the above procedure we finally obtain $\sigma_{8e} = 1.48$, $\sigma_{8S} = 0.68$, and $f_s = 0.79$, quite close to the desired values.

As another test of consistency with the real universe, the resultant auto-correlation functions of all the galaxies and of the E and S subsets are shown in Figure 9. They are to be compared to the observed $\xi(r) = (r/r_0)^{-\gamma}$, with $r_0 = 5.4$ h$^{-1}$Mpc and $\gamma = 1.8$ from the CfA redshift survey of optical galaxies (e.g. de Lapparent, Geller & Huchra 1988), and with $r_0 = 3.9$ h$^{-1}$Mpc and $\gamma = 1.57$ for *IRAS* 1.2Jy galaxies (Saunders *et al.* 1991).

The dispersion of 1D pair velocities for the S galaxies turns out to be $\sigma_v = 173$ km s$^{-1}$ at 4 h$^{-1}$Mpc, obeying the observational constraint specified in point 6 of §3.1.

*3.3 Assigning Observable Galaxy Properties*

The next step is to assign observable properties to each galaxy in order to mimic the selection into TF catalogs and the distance estimation as discussed in the following section.

We first truncate sharply the underlying galaxy distribution at $r = 120$ h$^{-1}$Mpc. Galaxies at larger distances are hardly relevant to the reconstruction that is currently applied inside 80 h$^{-1}$Mpc. This cutoff reduces the amount of work by more than 50% and it weakens the effects of the periodic boundary conditions. The cutoff introduces a Malmquist bias that can be easily corrected after the catalog is "observed" (§4.3 below).

The internal-rotation parameter, $\eta$ ($\equiv \log \Delta v - 2.5$; cf. WI), is first drawn for each galaxy at random from an assumed distribution function $\Phi(\eta)$, truncated at $\eta_{min}$. This cutoff is imposed by the finite mean number density of galaxies in the simulation such that $\int_{\eta_{min}}^{\infty} \Phi(\eta) d\eta = \bar{n}_g$. The $\eta$ function is obtained from an assumed Schechter luminosity function

$$\Phi(L) = \Phi_* L^{-\alpha} e^{-L}, \tag{15}$$

where $L$ is measured in units of $L_*$, with $M_* = -19.68$ and $\alpha = 1.07$ in the blue band. (Efstathiou, Ellis & Peterson 1988). The translation to an $\eta$ function is done via an assumed, tentative, deterministic, inverse TF relation,

$$\eta = a_I + b_I M. \tag{16}$$

For $\alpha = 1$, the $\eta$ function can be obtained analytically; it is roughly a step function truncated near $\eta^*$, corresponding to $M_*$. In practice, we draw an absolute magnitude at random from the Schechter function and translate it to $\eta$ via the assumed inverse TF relation with $a_I = -0.42$ and $b_I = -0.136$. This absolute magnitude is just a temporary device for the purpose of producing an appropriate unperturbed $\eta$ distribution.

We now assign absolute magnitudes according to an assumed forward TF relation,

$$M = a_f + b_f \eta + dM \tag{17}$$

where $dM$ is a random Gaussian variable with rms dispersion $\sigma_M$. The TF parameters $a_f$, $b_f$ and especially $\sigma_M$ are chosen to match those of the real data sets we try to mimic (Table 1). The specific way we scatter $M$ about a mean TF relation $M(\eta)$ (as opposed, for example, to scattering $\eta$) is crucial for the distances to be obtained properly by the assumed forward TF relation.

The apparent magnitude of each galaxy is finally computed via

$$m = M + 5 \log r + 25, \qquad (18)$$

with $r$ in units of h$^{-1}$Mpc. The inferred distance of each galaxy is then obtained from the "observed" $m$ and $\eta$ by

$$5 \log d = m - (a_f + b_f \eta) - 25. \qquad (19)$$

Measurement errors are neglected – the whole error is assumed to be due to scatter in the TF relation. The redshift of a galaxy of true velocity $\boldsymbol{v}$ at a true distance $r$ is $cz = r + \boldsymbol{v} \cdot \hat{\boldsymbol{r}}$, assuming no error in the measurement of redshift. The inferred radial peculiar velocity is $u = cz - d$.

## 4. MOCK MARK III CATALOGS

As a concrete useful example we create mock catalogs meant to mimic in certain detail the *Mark III* catalog of peculiar velocities (WI; WII; WIII), the sample currently used in the *POTENT* analyses (e.g. Dekel *et al.* 1996). The *Mark III* catalog consists of more than 3000 galaxies from several different data sets of S and E galaxies, calibrated and self-consistently put together as a homogeneous catalog for velocity analysis. The cluster data sets are treated in WI. The field galaxies are calibrated and grouped in order to minimize Malmquist biases in WII. The final catalog is tabulated in WIII, and will be distributed electronically.

In constructing the mock catalogs we first identify and select rich clusters, trying to mimic the true cluster data sets. The galaxies not associated with these clusters are then candidates for successive selection into mock field samples, following the selection procedure in each of the true field data sets. The field galaxies are grouped by the grouping algorithm of WII, and then corrected for inhomogeneous Malmquist bias as in WIII and Dekel *et al.* (1996). The selection parameters for each of the data sets are listed in Table 1.

### 4.1. Rich Clusters

The *Mark III* catalog contains two whole-sky data sets of rich clusters (WI): 13 E clusters (Ecl, Faber *et al.* 1989), and 26 S clusters (HM, based on Mould *et al.* 1991; Han & Mould 1990; 1992), with several clusters common to the two sets. (The 26 S clusters are a subsample of the 32 spiral clusters discussed in WI; 6 clusters, which are embedded in volumes well-probed by the field spiral samples, were eliminated from POTENT analysis to avoid redundant sampling.)

The selection procedures for these cluster samples are not defined in exact terms. We assume that they are selected by richness and try to mimic this selection in the simulation as follows:

1. Identify the $\sim 100$ richest clusters in the volume-limited sample (§3) inside a sphere of radius 120 $h^{-1}$Mpc about the LG (see below), excluding an appropriate Galactic zone of avoidance for each of the two sets ($b$ in Table 1). The mean number density of these clusters is comparable to that of $R \geq 0$ Abell clusters.

2. Identify the cluster centers (see below), and assign to each cluster all the galaxies within a radius $R_c$. The values of $R_c$ were determined from the typical maximum radii of clusters in the true samples: $R_c = 4$ $h^{-1}$Mpc for E galaxies and $R_c = 6$ $h^{-1}$Mpc for S galaxies.

3. Select galaxies from the volume-limited sample of clusters according to the apparent magnitude limits of the true samples ($m_b$ in Table 1), and define the "apparent" cluster richness accordingly. Compute the mean cluster distance, TF inferred distance (see below), and redshift.

4. Apply heliocentric redshift limits to the cluster samples ($cz_{max}$ in Table 1), and keep the 26 richest clusters of S galaxies and the 13 richest clusters of E galaxies, based on their apparent richness.

5. Reduce the total number of galaxies in clusters with equal probability per galaxy, such that it matches the number in the true sample.

The cluster finding in steps (1) and (2) was done in two steps because of computing limitations. First, cluster candidates were found by a friends-of-friends percolation algorithm, which was applied to a random subset of one in ten particles from the simulation. A maximum neighbor separation of 2.5 $h^{-1}$Mpc was used, corresponding to a density contrast of $\sim 25$ near the cluster edges. Only clusters with 4 members or more in the reduced subset were allowed. Tentative centers of mass were defined from the reduced clusters. Then, the cluster centers were re-defined by maximizing the number of particles within a sphere of 1.5 $h^{-1}$Mpc radius about them, considering all the simulated particles within 3 $h^{-1}$Mpc radius about each tentative center. Figure 10 shows the projected locations of the volume-limited sample of $R \geq 0$ clusters from a slice of thickness $\pm 10$ h$^{-1}$Mpc about the Supergalactic plane as found by applying this procedure.

### 4.2. Field Galaxies

The *Mark III* catalog includes four main data sets of field S galaxies (WII) as follows:

1. A82; a whole-sky, nearby sample by Aaronson *et al.* (1982), as re-calibrated by Tormen & Burstein (1995).

2. W91; a deep sample limited to the Perseus-Pisces region by Willick (1991).

3. CF; a northern-sky sample by Courteau & Faber (Courteau 1992).

4. MAT; a southern-sky sample by Mathewson *et al.* (1992) containing more than 1000 galaxies.

This sample of field S galaxies has been combined in the *Mark III* catalog with the E sample of the older *Mark II* catalog (Burstein 1990), which was mainly based on the survey by Lynden-Bell *et al.* (1988).

The mock data sets are selected one by one in the above order (E first), such that a given galaxy is allowed to be selected into one data set only. Cluster galaxies (§4.1.) are excluded from the field samples. Each mock data set is confined to the appropriate geometrical angular boundaries and heliocentric redshift cutoffs (Table 1).

For each data set we apply effective blue-band bright ($m_{min}$) and faint ($m_{max}$) magnitude limits, trying to approximate the observational procedure. In the real data sets (WI, WII) the observational selection criteria were relatively complicated. The selection variable (photographic magnitude or diameter) differed from the apparent magnitude on which TF distance estimation was based, and there was scatter in the correlation between the two. To straightforwardly mimic these effects in the mock data, we smear the magnitude limits: both $m_{min}$ and $m_{max}$ underwent Gaussian scattering about their means with standard deviation $dm$ (Table 1); all three parameters were specific to each data set.

For the purpose of applying the magnitude limits, the magnitudes are subjected to Galactic extinction as a function of Galactic latitude (Fisher & Tully 1981). The magnitudes kept for distance estimation are the uncontaminated magnitudes, to match the observed magnitudes after correction for extinction. (Note that this procedure assumes that the Galactic extinction corrections are "perfect," which they may not be in the real catalog.)

Finally, the number of galaxies in each mock data set is cut down at random to match the number of galaxies in the true data set. This mimics the exclusion of galaxies based on properties other than magnitudes, such as inclination, galaxy sub-type, etc. The reduction factor typically ranges from 1.6 to 3. In the MAT $\delta > -17.5$ subsample the reduction factor is $\sim 10$ due to the incompleteness of the optical catalog (MCG) from which this sample was drawn. In the CF sample the reduction factor is as high as $\sim 50$ due to a very conservative selection criteria for inclination and morphology.

The success of the above procedure is tested by comparing the number of galaxies in each data set as functions of redshift and of magnitude, $N(z)$ and $N(m)$, to the corresponding observed distributions (Figure 12). In this comparison, the blue magnitudes are shifted by $m_{shift}$ (Table 1), to mimic the transformation from the blue-band to the actual filter used in each specific observation. The fits between the mock and observed $N(z)$ and $N(m)$ are fine-tuned by allowing small adjustments in $m_b$ and $dm$.

### 4.3. Grouping and Correcting for Malmquist Bias

The random scatter in the distance estimator is a source of systematic biases in the inferred distances and peculiar velocities, which are generally termed "Malmquist" biases (*e.g.* Lynden-Bell *et al.* 1988; Willick 1994). One bias is in the *calibration* of the forward TF relation because of the magnitude limit in the selection. Another bias is in the *inferred distance*, $d$, and the associated mean peculiar velocity at a given $d$. In this case, the combination of distance errors and galaxy-density variations along the line of sight systematically enhances the inferred velocity gradients and thus the inferred density fluctuations. This is the *inhomogeneous* Malmquist bias (IM). These biases are treated in the *Mark III* catalog and in the forward POTENT analysis in two steps: first grouping, which simultaneously minimizes the calibration bias and reduces the IM bias, and second a systematic correction for IM bias. In order to test these procedures we repeat them in the mock catalog.

The Mark III spiral samples were analyzed using a grouping algorithm described in detail in § 2.2.2 of WII. Briefly, this algorithm first links objects that pass redshift-space proximity tests, and subsequently checks whether objects thus grouped are reasonably close in TF-distance space as well. The redshift-space proximity requirement is enforced much more strongly than the TF-distance one. The rms redshift-space group size is typically $\lesssim 200$ km s$^{-1}$ in the radial direction, and about twice that in the transverse direction, whereas the TF-distances of group memberse can differ fractionally by up to $\sim 3\Delta$ (where $\Delta$ is the relative distance error for the object), or up to several thousand km s$^{-1}$ in some cases. The TF-distance criterion is applied only to minimize the grouping of objects whose peculiar velocities cause them to coincide in redshift, but that are in reality widely separated along the line of sight. For the real data, preliminary TF relations derived from Hubble flow fits were used in the grouping (WII), whereas for the simulated samples the true TF relations were used. The grouping reduces the IM bias by dividing the distance error of each group of $N$ members by a factor of $\sqrt{N}$.

In order to correct the grouped data for IM bias, the noisy inferred distance of each object, $d$, is replaced by the expectation value of the true distance, $r$, given $d$ (Willick 1991, eq. 5.70):

$$E(r|d) = \frac{\int_0^\infty r^3 n(r) \exp\left(-\frac{[\ln(r/d)]^2}{2\Delta^2}\right) dr}{\int_0^\infty r^2 n(r) \exp\left(-\frac{[\ln(r/d)]^2}{2\Delta^2}\right) dr} , \qquad (20)$$

where $\Delta \simeq 0.46\sigma_{\rm TF}$. For single galaxies, $n(r)$ is the number density of galaxies in the underlying distribution of galaxies from which galaxies were selected for the sample (by quantities that do not explicitly depend on $r$). We use here the G5 smoothed density field of the galaxies in the simulation itself (Fig. 8), to be approximated when applied to the real data by the density field of *IRAS* galaxies, for example. The artificial cutoff of the galaxy distribution at $r = 120$ h$^{-1}$Mpc is fully corrected for by setting $n(r) = 0$ beyond that distance. The effect of redshift limits in the different data sets are corrected in a similar way, assuming that at large distances a cutoff in redshift in the CMB frame is a reasonable approximation to a cutoff in true distance. In grouped data, the density run $n(r)$ is multiplied by an appropriate grouping correction factor (Dekel *et al.* 1996).

### 4.4. Artificial IRAS Redshift Surveys

It is straightforward to construct magnitude-limited redshift surveys from the simulated galaxy distribution. These catalogs can serve for testing reconstruction methods from redshift surveys, and for testing comparisons of redshift data and velocity data aimed at estimating the cosmological density parameter $\Omega$ and studying how galaxies trace mass.

We have produced so far several random mock catalogs of the *IRAS* 1.2Jy redshift survey by simply applying its radial selection function to the S sample (Yahil *et al.* 1991, Fisher 1992):

$$\phi(r) = \left(\frac{r}{r_s}\right)^{-2\alpha} \left(\frac{r^2 + r_*^2}{r_s^2 + r_*^2}\right)^{-\beta} , \qquad (21)$$

with $r_s = 500$ km s$^{-1}$, $r_* = 5184$ km s$^{-1}$, $\alpha = 0.492$ and $\beta = 1.830$. Figure 13 shows the similarity between $N(z)$ in one of the mock catalogs and in the true data.

### 5. CONCLUSION

We have described a multi-stage procedure for constructing N-body simulations that mimic our cosmological neighborhood, and for selecting from them mock catalogs of galaxies and clusters that resemble the main data set currently used in dynamical studies of large-scale structure.

The primary motivation for this effort was to create a reliable tool for quality control of *POTENT* reconstruction methods. For this purpose we generated mock catalogs that resemble in detail the *Mark III* catalog of peculiar velocities. It is essential that the underlying mass distribution and velocity field, the volume-limited galaxy distribution, the observable properties of galaxies, the sampling, the grouping, and the rest of the observational procedures all mimic as closely as possible the true universe, because the various systematic errors depend sensitively on these different aspects of the data and on the correlations between them.

We have succeeded in reproducing the large-scale features in reasonable detail and the small-scale properties in a more statistical sense. This is demonstrated by the similarity of the smoothed density fields in the simulation and in the input *IRAS* galaxy distribution, and by the resemblance of the distribution of galaxies as functions of space, redshift and magnitudes.

However, the current mock catalogs have several limitations. The sparseness of sampling by the *IRAS* 1.2 Jy galaxies limits the input resolution possible in principle, as a function of distance. The resolution of the constraining density field is therefore limited to $\sim 5$ $h^{-1}$Mpc within the volume currently covered by velocity data. A similar limitation is imposed by our quasi-linear methods for recovering the real-space distribution of *IRAS* galaxies and tracing it back in time to the linear regime. This resolution is marginally suitable for reproducing the rich clusters roughly where they are in the real universe, but it is not sufficient for recovering the true galaxies and groups of galaxies at their true positions. The method of constrained realizations maximizes the signal from the smoothed input data, and it produces small-scale structure that mimics the true structure as well as possible in a statistical sense subject to the observed constraints.

The resolution is also limited by our current usage of a PM N-body code, with spatial and mass resolution of only $\sim 2$ $h^{-1}$Mpc. The identification of galaxies with individual particles of the simulation is an approximation which may overlook possible velocity biasing due to internal galactic degrees of freedom, and it limits the density biasing to simplified schemes. Simulations of higher resolution will be relatively straightforward to carry out in the near future.

Although we have applied a reasonable biasing scheme, it remains somewhat ad-hoc and over-simplified. We introduced some scale dependence and some non-linear density dependence, but this is only one possibility out of many. Perhaps even more importantly, we have ignored the statistical nature of the biasing process, which may alter the results (e.g. Dekel & Lahav, in preparation). An immediate area for improvement is the investigation of the reconstruction methods under several different biasing schemes.

The simulated *Mark III* and *IRAS* catalogs we have produced are already serving the *POTENT* team in the testing and development of improved reconstruction methods. We apply the methods to the mock catalogs and compare the recovered dynamical fields with the "true" fields smoothed directly from the particles of the simulation. This comparison allows us to measure the systematic and random errors, and to improve the methods accordingly. We propose that this set of Monte Carlo catalogs will serve the community of practitioners of reconstruction methods as a standard benchmark. For this purpose we will make the mock catalogs available electronically upon request, in parallel with the *Mark III* catalog itself (WIII).

Furthermore, the method described above, either as a whole or component by component, can serve to construct mock catalogs of similar nature that will mimic new data as they become available. We will be pleased to help others to construct mock catalogs based on any new data.

We thank Amos Yahil and Michael Strauss for providing the *IRAS* density field, and Adi Nusser for the time-machine code. This work was supported in part by the US-Israel Binational Science Foundation grant 92-00355, the Israel Science Foundation grant 462/92, and the US National Science Foundation grant PHY-91-06678.

## APPENDIX A: SIMULATING THE MORPHOLOGY-DENSITY RELATION

We wish to translate a number density, $n_g$, derived from the volume-weighted sample of galaxies drawn from the simulations, into the number density $n_d$ as used by Dressler (1980) in his morphology-density relation, which we approximate by Eq. (12).

The challenge is to obtain $f_{pm}$ (Eq. 14), to plug into Eq. (13). Figure 14 shows the cumulative count of galaxies, $N(>n)$ as a function of $n$, referring to densities in Dressler's language. The solid circles represent the actual distribution of Dressler as read from the top of his Figure 4. The triangles represent the distribution of $n_{70}$ (see below) in the simulation, after biasing $b = 1.35$. The desired $n_d(n_g)$ will be obtained by matching the corresponding distributions of Dressler and of the simulation galaxies, after accounting for the built-in differences between them. They differ in several ways, and we match them step by step as follows:

*Step 1: Adjusting the Mean Densities*

The mean densities differ because of different absolute-magnitude limits. Define $\beta \equiv \bar{n}_d/\bar{n}_g$. We find (below) $\beta = 0.145$. Then $n_g$ is multiplied by $\beta$ to match $n_d$. The horizontal shift by $\log \beta$ yields the line of squares in Fig. 13. The density $n_{10}$ of Dressler should in turn be replaced in the simulation by $n_{10/\beta}$, i.e. $n_{70}$.

To evaluate $\beta$, we first need to compute $\bar{n}_d$, which would have been Dressler's mean density had he sampled all regions of space uniformly, independent of density, rather than focusing on cluster regions. Dressler corrected $n_d$ such that it reflects galaxies of

$M_v < -20.4$ using $h = 0.5$. This corresponds to $M_v < -18.895 + 5\log h$, which corresponds to

$$L_{lim} = 10^{0.4(M_\odot - M_{lim})} L_\odot = 3.09 \times 10^9 h^{-2} L_\odot \qquad (A1)$$

given that $M_{\odot v} = 4.83$.

We adopt Schechter's luminosity function (Eq. 15), $\phi(L) = \phi_* L^{-\alpha} e^{-L}$, with $\alpha = 1.07$, $\phi_* = 0.010\,(\text{h}^{-1}\text{Mpc})^{-3}$, and $L$ measured in units of $L_*$. Since we adopt $M_{*B} = -19.68 + 5\log h$, and $M_{\odot B} = 5.48$, we get

$$L_* = 10^{0.4(M_\odot - M_*)} L_\odot = 1.16 \times 10^{10} h^{-2} L_\odot. \qquad (A2)$$

Thus, $L_{lim}$ in units of $L_*$ is 0.267. Finally,

$$\bar{n}_d = \phi_* \int_{L_{lim}}^\infty L^{-\alpha} e^{-L} dL = 1.027 \phi_* = 0.010(\text{h}^{-1}\text{Mpc})^{-3}. \qquad (A3)$$

Since $\bar{n}_g = 0.071$, we obtain $\beta \equiv \bar{n}_d / \bar{n}_g = 0.145$.

*Step 2: Adjusting the Volume Sampled*

The difference in the total volume sampled affects linearly the total number count. This corresponds to a vertical shift in the log-log plot of Figure 14 by a multiplicative factor $\alpha$. If Dressler had sampled uniformly, then we could have corrected for this difference by the values of $N(n > 0)$. However, Dressler's sampling is strongly biased against low densities. This shows as strong flattening of $N(> n)$ (solid circles) for $n < 200$, and it does not enable a straightforward determination of $\alpha$.

To make things worse, $n_g$ is severely underestimated at the high end because of the limited grid *resolution* of the PM code. This is noticed in the distribution (squares) for $n > 20$ (which is indeed roughly where it is expected to be based on 70/cell − volume ~ 10].

Fortunately, the sampling of Dressler seems uniform in the range $200 < n < 20000$ (i.e. within clusters), as indicated by the constant logarithmic slope of the distribution there, $\nu \simeq -0.63$. The simulation distribution has a good part too: it is not affected by the PM resolution for $n < 10$, and it's logarithmic slope there is also $\nu \simeq -0.63$. These uncontaminated parts of the curves allow us to determine $\alpha$ by shifting the simulated distribution (suares) upwards until the lines of constant slope in the two distributions become a natural extension of each other. With the sub-volume used in determining the simulated distribution, we find $\alpha = 6$. This yields the line of pentagons in Fig. 13.

*Step 3: Deriving $f_{pm}$*

The combined line of hexagons in Figure 14, made of the pentagons at small densities and merging smoothly into the solid circles at large density, is the corrected distribution of Dressler, had he sampled regions of all densities uniformly. The desired relation between $n_g$ and $n_d$ can be read from the figure by comparing the lines of hexagons and pentagons at the same $N$ values. Equation (14) is a functional fit. In fact, the line of hexagons in the figure is derived from the line of pentagons by Eq. (14).

# FIGURE CAPTIONS

**Figure 1 :** The present-day density fluctuation field as recovered from the *IRAS* 1.2Jy redshift survey via the power-preserving filter, smoothed with a Gaussian of radius 5 $h^{-1}$Mpc (G5). The mean, $\delta = 0$, is marked by the heavy contour, while positive and negative density fluctuations are marked by solid and dashed contours respectively, with contours spacing $\Delta\delta = 0.2$.

**Figure 2 :** The G5 density fluctuation field after being traced back to the linear regime. The normalization is arbitrarily $\sigma_8 = 1$. Contours are as in Figure 1.

**Figure 3 :** The probability distribution functions of initial G5 density fluctuations in sub-volumes ($R < 7000$ km s$^{-1}$, $7000 < R < 10000$ km s$^{-1}$) and in the whole sphere ($R < 12800$ km s$^{-1}$), before and after Gaussianization ((a) and (b) respectively)

**Figure 4 :** The G5, linear density fluctuation field after Gaussianization. The normalization is arbitrarily $\sigma_8 = 1$. Contours are as in Figure 1.

**Figure 5 :** The initial density fluctuation field including a constrained realization of small-scale power smoothed with a Gaussian of radius 2 $h^{-1}$Mpc (G2). The normalization is arbitrarily $\sigma_8 = 1$. Contours are as in Figure 1.

**Figure 6 :** The projected mass distribution at the final time of the simulation in a slice of thickness $\pm 10$ $h^{-1}$Mpc about the Supergalactic plane. The corresponding density fluctuation field, smoothed with a Gaussian of radius 12 $h^{-1}$Mpc (G12), is mapped by the contours as in Figure 1.

**Figure 7 :** A map of the G5 mass density fluctuation field at the final time of the simulation in the Supergalactic plane. Contours are as in Figure 1.

**Figure 8 :** The G5 density fluctuation field of E galaxies (left) and S galaxies (right). Contours are spaced here by $\Delta\delta = 0.5$ (different from the previous figures).

**Figure 9 :** Two-point auto-correlation functions of galaxies: all (A), E's, and S's. The reference dotted line is for $r_0 = 5 \ h^{-1}$Mpc and $\gamma = 1.8$.

**Figure 10 :** All rich ($R \geq 0$) cluster locations (bold circles) in a slice of thickness $\pm 10 \ h^{-1}$Mpc about the SG plane. The clusters are shown on top of the particles distribution at the final stage of the simulation.

**Figure 11 :** The projected distribution of galaxy inferred positions (not corrected for biases) in a slice of thickness $\pm 20 \ h^{-1}$Mpc about the Supergalactic plane. Left: real data. Right: mock data. The G12 mass fluctuation field is indicated by the contours of spacing $\Delta\delta = 0.2$ at the background.

**Figure 12 :** The distribution functions of redshifts and apparent magnitudes for the galaxies in the real data (heavy lines) and in the mock data (thin lines)

**Figure 13 :** The redshift distribution functions in the *IRAS* 1.2Jy survey: real (bold) versus mock (thin).

**Figure 14 :** Matching the probability distribution of $n_{70}$ densities in the simulation (open triangles) to that observed by Dressler (solid circles). The dotted-dashed line connecting hexagons is the corrected distribution of Dressler, had he sampled regions of all densities uniformly. See details in Appendix A.

**Table 1 :** Selection parameters for the Mark III mock data. **1.** Abbreviation for the data set. **2.** Redshift limit (Helio-centric). **3, 4.** Geometrical boundaries in celestial declination ($\delta$) and right ascension ($\alpha$), in degrees. **5.** Geoemtrical boundary in Galactic latitude ($b$). **6.** Mean apparent magnitude limit (blue). **7.** Scatter of the magnitude limit. **8.** Additive correction from the blue magnitude of the mock catalog to the actual observed band. **9,10.** Forward Tully-Fisher parameters. **11.** Scatter in the TF relation in terms of absolute magnitude at a given $\eta$. **12.** Number of galaxies in the real data set. The number in the mock data set may differ by up to 10%. **13.** Number of clusters in the data set.